\documentclass[aps,prb,twocolumn,showpacs,showkeys,preprintnumbers,groupedaddress,amsmath,amssymb,]{revtex4-1}

\usepackage{graphicx}
\usepackage{dcolumn}
\usepackage{bm}
\usepackage[dvips]{color}

\begin{document}




\title{Simultaneous observation of small- and large-energy-transfer electron-electron scattering in three dimensional indium oxide thick films}


\author{Yang Yang}
\author{Xin-Dian Liu}
\author{Zhi-Qing Li}
\email[Author to whom correspondence should be addressed. Electronic mail: ]{zhiqingli@tju.edu.cn}

\affiliation{Tianjin Key Laboratory of Low Dimensional Materials Physics and
Preparing Technology, Department of Physics, Tianjin University, Tianjin 300072,
China}


\date{\today}

\begin{abstract}
In three dimensional (3D) disordered metals, the electron-phonon (\emph{e}-ph) scattering is the sole significant inelastic process. Thus the theoretical predication concerning the electron-electron (\emph{e}-\emph{e}) scattering rate $1/\tau_\varphi$ as a function of temperature $T$ in 3D disordered metal has not been fully tested thus far, though it was proposed 40 years ago [A. Schmid, Z. Phys. \textbf{271}, 251 (1974)].
We report here the simultaneous observation of small- and large-energy-transfer \emph{e}-\emph{e} scattering in 3D indium oxide thick films. In temperature region of $T\gtrsim100$\,K, the temperature dependence of resistivities curves of the films obey Bloch-Gr\"{u}neisen law, indicating the films possess degenerate semiconductor characteristics in electrical transport property. In the low temperature regime,  $1/\tau_\varphi$ as a function of $T$ for each film can not be ascribed to \emph{e}-ph scattering.  To quantitatively describe the temperature behavior of $1/\tau_\varphi$, both the 3D small- and large-energy-transfer \emph{e}-\emph{e} scattering processes should be considered (The small- and large-energy-transfer \emph{e}-\emph{e} scattering rates are proportional to $T^{3/2}$ and $T^2$, respectively). In addition, the experimental prefactors of $T^{3/2}$ and $T^{2}$ are proportional to $k_F^{-5/2}\ell^{-3/2}$ and $E_F^{-1}$ ($k_F$ is the Fermi wave number, $\ell$ is the electron elastic mean free path, and $E_F$ is the Fermi energy), respectively,  which are completely consistent with the theoretical predications. Our experimental results fully demonstrate the validity of theoretical predications concerning both small- and large-energy-transfer \emph{e}-\emph{e} scattering rates.
\end{abstract}


\maketitle

\section{Introduction}
Four decades ago, the inelastic scattering of the electrons of an impure metal by a screened Coulomb
interaction had been investigated.\cite{ref1} It had been found that the electron-electron (\emph{e}-\emph{e}) scattering rate $1/\tau_{ee}$ in three dimensional (3D) disordered metal can be expressed as,\cite{ref1,ref2,ref3}
\begin{equation}\label{Eq.(EE)}
\frac{1}{\tau_{ee}}=\frac{\pi}{8}\frac{(k_{B}T)^{2}}{{\hbar}E_F}+\frac{\sqrt3}{2{\hbar}\sqrt{E_F}}\left(\frac{k_{B}T}{k_{F}\ell}\right)^{3/2},
\end{equation}
where $E_F$ is Fermi energy, $k_F$ is the Fermi wave number, $\ell$ is the mean free path of electrons, $T$ is the temperature, $k_B$ is the Boltzmann constant and $\hbar$ is the Planck constant divided by 2$\pi$. The first and second terms on the right hand side of Eq.~(\ref{Eq.(EE)})  represent the large- and small-energy-transfer \emph{e}-\emph{e} scattering processes and dominate at energy scale $\varepsilon>\hbar/\tau_e$ and $\varepsilon<\hbar/\tau_e$, respectively, where $\tau_e$ is the electron elastic mean free time.\cite{ref2,ref3} However, the validity of Eq.~(\ref{Eq.(EE)}) has not been completely tested\cite{ref4,ref5,ref6,ref7,ref8} due to the \emph{e}-\emph{e} scattering is
negligible compared with electron-phonon (\emph{e}-ph) scattering in general
3D disordered metals.\cite{ref9,ref10,ref11} Taking advantage of the low concentration and free-electron-like electrical transport properties of charge carrier in Sn doped In$_2$O$_3$ (ITO),\cite{ref12,ref13} Zhang \emph{et al} recently tested the correctness of the small-energy-transfer term in Eq.~(\ref{Eq.(EE)}) in ITO thick films.\cite{ref14} In the present paper, we fully demonstrate the validity of Eq.~(\ref{Eq.(EE)}) (including both large- and small-energy-transfer terms) in In$_2$O$_3$ thick films. Our results concerning the electrical transport properties and dephasing mechanism of this low carrier concentration and large $k_F\ell$ material are presented and discussed below.

\begin{figure}
\begin{center}
\includegraphics[scale=1]{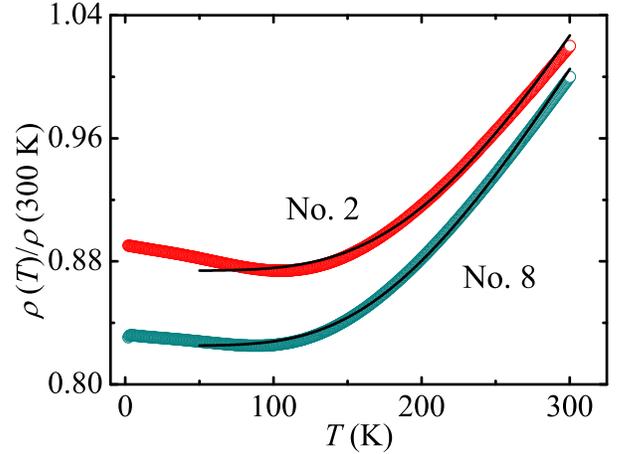}
\caption{(Color online) Normalized resistivity as a function of temperature for films Nos. 2 and 8, films. The solid curves are least-squares fits to Eq.~(\ref{Eq.(RT)}). For clarify, the data for film No. 2 has been shifted by +0.02.}
\label{FIGResistivity}
\end{center}
\end{figure}

\section{Experimental methods}
In$_2$O$_3$ films were deposited on the (100) yttrium stabilized ZrO$_2$ (YSZ) single crystal substrates by standard rf-sputtering method. An In$_2$O$_3$ target with purity of 99.99\% and diameter of 60 mm was used as the sputtering source. The base pressure of chamber was less than $1\times10^{-4}$ Pa, and the sputtering  was carried out in argon atmosphere with pressure of 0.55\,Pa. In the depositing process, the substrate temperature was kept at 703, 723, 743, 763, 803, 823, 843, and 923\,K, respectively, to tune the charge concentration and the disorder degree of the film. Hall-bar-shaped samples (1-mm wide and 3.6-mm long), deposited by using metal masks,\cite{ref14,ref15} were used to measure the electrical transport properties.  The thicknesses of the films ranging from $\sim$950\,nm to $\sim$1100\,nm were measured using a surface profiler (Dektak, 6M). The structures of the films were determined by x-ray diffraction (XRD) with Cu$K_\alpha$ radiation. The measurements include normal $\theta$-$2\theta$, $\phi$, and $\omega$ scans. The XRD results indicate that the films are epitaxially grown on (100) YSZ substrates along [100] direction. The Hall coefficient, longitudinal resistivity and low-field magnetoconductivity  were measured on a physical property measurement system (PPMS-6000, Quantum Design) by four-probe method. In the magnetoconductivity measurements, the magnetic field was applied perpendicular to the film plane.

\begin{table*}
\caption{\label{TableLi} Sample parameters for the eight 3D In$_2$O$_3$ films. $T_s$ is the substrate temperature during deposition, $d$ is the mean film
thickness, $\rho$ is the resistivity, $n$ is the carrier concentration, and $\theta_{D}$ is the Debye temperature. $1/\tau_{0}$, $A^S_{ee}$, and $A^L_{ee}$ are defined in Eq.~(\ref{Eq.(EE)}). $(A^S_{ee})^{th}$ and $(A^L_{ee})^{th}$ are the theoretical values of $A^S_{ee}$, and $A^L_{ee}$ predicted by Eq.~(\ref{Eq.(EE)}).}
\begin{ruledtabular}
\begin{center}
\begin{tabular}{ccccccccccccccc}
Film & $T_s$ & $d$ & $\rho$(300\,K) & $\theta_{D}$ & $k_{F}l$ & $n$(10\,K) & $1/\tau^{0}_{\varphi}$ & $A^S_{ee}$ & $A^L_{ee}$
& $(A^S_{ee})^{th}$ &  $(A^L_{ee})^{th}$ \\
 & (K) & (nm) & (m$\Omega$\,cm) & (K) & & ($10^{19}$\,cm$^{-3}$) & ($10^{-9}$\,s$^{-1}$) & (K$^{-3/2}$\,s$^{-1}$)
 & (K$^{-2}$\,s$^{-1}$) & (K$^{-3/2}$\,s$^{-1}$) & (K$^{-2}$\,s$^{-1}$) & \\  \hline

1 & 703 & 1093.60 &1.52 & 1456 &8.9 &3.83 &0.88 & 1.28$\times10^{8}$ & 3.02$\times10^{7}$ & 1.23$\times10^{8}$  & 4.27$\times10^{7}$\\
2 & 723 & 1068.36 &2.50 & 1399 &5.3 &4.06 &14.2 & 1.69$\times10^{8}$ & 2.81$\times10^{7}$ & 2.64$\times10^{8}$  & 4.11$\times10^{7}$\\
3 & 743 & 1050.98 &2.55 & 1399 &5.1 &3.91 &1.74 & 1.61$\times10^{8}$ & 2.67$\times10^{7}$ & 2.79$\times10^{8}$  & 4.21$\times10^{7}$\\
4 & 763 & 1067.45 &2.37 & 1373 &5.6 &4.26 &23.9 & 1.52$\times10^{8}$ & 2.54$\times10^{7}$ & 2.38$\times10^{8}$  & 3.97$\times10^{7}$\\
5 & 803 & 1033.56 &1.89 & 1365 &6.7 &4.92 &16.6 & 1.37$\times10^{8}$ & 2.43$\times10^{7}$  & 1.72$\times10^{8}$  & 3.61$\times10^{7}$\\
6 & 823 & 1002.18 &1.67 & 1313 &7.9 &4.71 &20.9 & 1.20$\times10^{8}$ & 2.32$\times10^{7}$ & 1.37$\times10^{8}$  & 3.72$\times10^{7}$\\
7 & 843 & 1016.62 &1.50 & 1335 &8.5 &5.22 &1.84 & 1.20$\times10^{8}$ & 2.24$\times10^{7}$ & 1.20$\times10^{8}$  & 3.47$\times10^{7}$\\
8 & 923 & 952.93 &1.31 & 1236 &9.1 &6.31 &2.43 & 1.15$\times10^{8}$ & 1.87$\times10^{7}$ & 1.01$\times10^{8}$  & 3.06$\times10^{7}$\\
\end{tabular}
\end{center}
\end{ruledtabular}
\end{table*}

\section{Results and Discussion}
Figure~\ref{FIGResistivity} shows the resistivity $\rho$ as a function of temperature $T$ for two representative In$_2$O$_3$ films, as indicated. The resistivities decrease with decreasing temperatures from 300  to $\sim$100\,K, reach their minimum and then increase with further decreasing temperature down to our minimum measuring temperature, 2\,K. We compare the $\rho(T)$ data at higher temperature region with Bloch-Gr\"{u}neisen law\cite{ref16}
\begin{equation}\label{Eq.(RT)}
\rho=\rho_0+\beta{T}\left(\frac{T}{\theta_D}\right)^4\int^{\theta_{D}/T}_{0}\frac{x^{5}dx}{(e^x-1)(1-e^{-x})},
\end{equation}
where $\rho_0$ is the residual resistivity, $\beta$ is a constant, $\theta_{D}$ is the Debye temperature. The solid curves in Fig.~\ref{FIGResistivity} are least-squares fits to Eq.~(\ref{Eq.(RT)}). The resistivity data can be well described by Eq.~(\ref{Eq.(RT)}), indicating In$_2$O$_3$ films possess highly degenerate semiconductor characteristic in electrical transport properties. The fitted values of $\theta_{D}$ are listed in Table~\ref{TableLi}. Similar to Sn doped In$_2$O$_3$\cite{ref12} and F doped SnO$_2$,\cite{ref17} our In$_2$O$_3$ films also possess higher Debye temperature. The enhancement in resistivity with decreasing temperature below $\sim$100\,K can be attributed to the weak-localization (WL) and electron-electron interaction effects,\cite{ref18,ref19,ref20,ref21,ref22} which is similar to that in ITO\cite{ref8,ref12} and FTO.\cite{ref17} The temperature behavior of resistivity of the In$_2$O$_3$ films in higher temperature regime indicate In$_2$O$_3$ possesses degenerate semiconductor characteristic in transport properties. In fact, the origins of the high conductivity and degenerate semiconductor characteristic of In$_2$O$_3$ are still enigmatic. The oxygen vacancy is generally considered as the main contribution to the high conductivity of the undoped In$_2$O$_3$.\cite{ref23,ref24,ref25,ref26,ref27,ref28,ref29,ref30} However, recent theoretical results indicate that the donor level of oxygen vacancies is too deep to produce large densities of free electrons at room temperature.\cite{ref31} The reasons that In$_2$O$_3$  possesses relative high carrier concentration and behaves as degenerate semiconductor in electrical transport properties deserve further investigations.

\begin{figure}
\begin{center}
\includegraphics[scale=1]{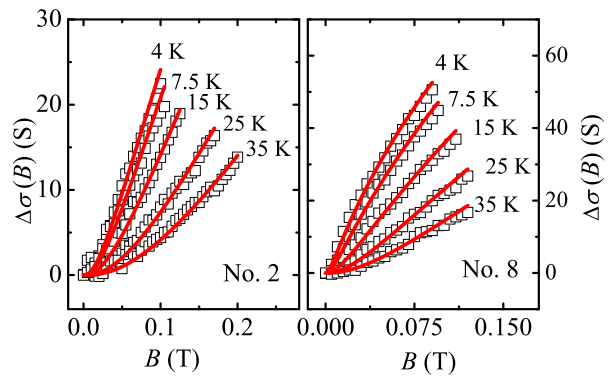}
\caption{(Color online) Magnetoconductivity versus magnetic field for films Nos. 2 and 8 measured at different temperatures. The magnetic field was applied perpendicular to the film plane. The solid curves are least-squares fits to Eq.~(\ref{Eq.(MC)}).}
\label{FIGMC}
\end{center}
\end{figure}

Figure~\ref{FIGMC} shows the magnetoconductivity, $\Delta\sigma=\sigma(B)-\sigma(0)$, as a function of magnetic field at low temperatures (4-35 K) for two representative films, as indicated. We found that the magnetoconductivity is positive and its magnitude at a certain field decrease with increasing temperature. These features indicate that the spin-orbit scattering is weak, and the WL effect governs the behaviors of $\sigma(B)$ at low field region.\cite{ref9,ref21,ref22} Considering the thicknesses of all samples are $\sim$1\,$\mu$m, we analyse the magnetoconductivity data using 3D WL theory. In 3D disordered conductors, the magnetoconductivity  due to WL effect is given by\cite{ref32,ref33,ref34,ref35,ref36}
\begin{equation}\label{Eq.(MC)}
\Delta\sigma=\sigma(B)-\sigma(0)=\frac{e^2}{2\pi^{2}\hbar}\sqrt{\frac{eB}{\hbar}}f_{3}\left(\frac{B}{B_{\varphi}}\right),
\end{equation}
where $e$ is the elementary charge, $D$ is the diffusion constant, and $B_\varphi=B_0+B_{\rm i}$. The characteristic field  $B_j$ is defined by $B_j=\hbar/(4eD\tau_j)$, where $j=0$ and i represent the $T$-independent and inelastic scattering fields ($\tau_j$ is the corresponding relaxation time), respectively. The theoretical predictions of Eq.~(\ref{Eq.(MC)}) are least squares fitted to our magnetoconductivity data and are shown as solid curves in Fig.~\ref{FIGMC}. Our magnetoconductivity data can be well described by Eq.~(\ref{Eq.(MC)}). For the samples, the obtained electron dephasing length  $L_\varphi=\sqrt{D\tau_\varphi}=\sqrt{\hbar/(4eB_\varphi)}$ at 4 K varies from $\sim$90 to $\sim$380 nm, which much less than the thicknesses of the films. Hence our films are 3D with regard to WL effect.

\begin{figure}[htp]
\begin{center}
\includegraphics[scale=1]{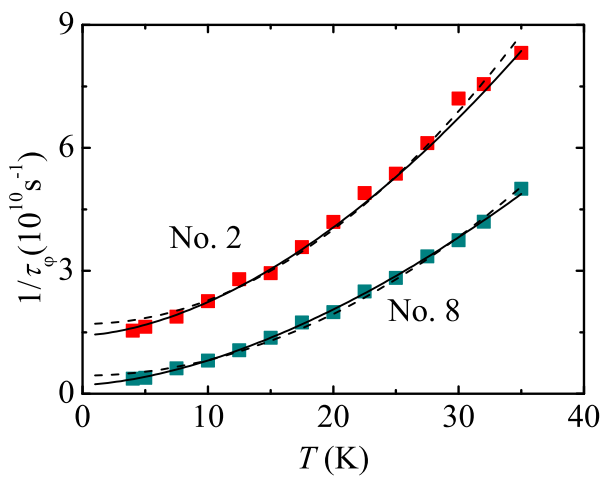}
\caption{(Color online) The electron dephasing rate as a function of temperature for films Nos. 2 and 8. The dash curves are least-squares fits to Eq.~(\ref{Eq.(EPfit)}), and the solid curves are least-squares fits to Eq.~(\ref{Eq.(EEfit)}).}
\label{FIGDephase}
\end{center}
\end{figure}

Figure~\ref{FIGDephase} shows the extracted electron dephasing rate as a function of $T$ for the two representative films, as indicated. As mentioned above, the \emph{e}-ph scattering is the dominant electron dephasing mechanism in  3D general disordered metals.\cite{ref9,ref10,ref11} Assuming the \emph{e}-ph scattering mechanism governs the electron dephasing processes of the films, we quantitatively analyze the $1/\tau_\varphi(T)$ data now. Theoretically, the electron scattering
by transverse vibrations of defects and impurities dominates the \emph{e}-ph relaxation. In the quasiballistic
limit of $q_T\ell > 1$ (where $q_T$ is the wave number of a
thermal phonon), the relaxation rate, $1/\tau_{e\text{-}t,\rm{ph}}$, is given by\cite{ref37,ref38}
\begin{equation}\label{Eq.(EP)}
\frac{1}{\tau_{e\text{-}t,{\text{ph}}}}=\frac{3\pi^{2}k^{2}_{B}\beta_{t}T^{2}}{(p_{F}u_{t})(p_{F}\ell)},
\end{equation}
where $\beta_{t}$=$(2E_{F}/3)^{2}N(E_F)/(2\rho_{m}u^2_t)$ is the electron-transverse phonon coupling constant, $p_F$ is the Fermi momentum, $u_t$ is the transverse sound velocity, $\rho_m$ is the mass density, and $N(E_F)$ is the electronic density of states at the Fermi level. For our films, the values of $q_T\ell$ vary from $\sim$$0.3T$ to $\sim$$0.5T$.\cite{note1}, which are greater than unity for $T\geq4$\,K.  Our $1/\tau_\varphi(T)$ data from 4 to 35\,K were least-squares fitted to the following equation:
\begin{equation}\label{Eq.(EPfit)}
\frac{1}{\tau}=\frac{1}{\tau_0}+\tilde{A}T^2,
\end{equation}
where $1/\tau_0$ stands for the saturated dephasing rate,\cite{ref39,ref40,ref41} and $\tilde{A}$ is an adjustable parameter and presumably represents the \emph{e}-ph scattering strength given in Eq.~(\ref{Eq.(EP)}). The dash curves in Fig.~\ref{FIGDephase} are the fitted results. Clearly, the experimental dephasing rate can be described by the $T^2$ term at high temperature regime ($T\gtrsim10$\,K), while it deviate from the predication of Eq.~(\ref{Eq.(EPfit)}) for $T<10$\,K. The fitted values of $\tilde{A}$ vary from $\sim$$3.8\times10^7$ to $\sim$$5.8\times10^7$\,K$^{-2}$\,s$^{-1}$. On the other hand, the values of $u_t$ and $\rho_m$ of In$_2$O$_3$ are 2400\,m\,s$^{-1}$ and 7100\,kg\,m$^{-3}$,\cite{ref42} respectively, the carrier concentrations vary from $3.8\times 10^{19}$ to $6.3\times 10^{19}$\,cm$^{-3}$, and the effective mass of electron $m^\ast$ can be taken as $m^\ast=0.4m_e$,\cite{ref43} where $m_e$ is the free-electron mass.
According to Eq.~(\ref{Eq.(EP)}), one can readily obtain that the theoretical values of \emph{e}-ph scattering strength $A_{e\text{-}t,\rm{ph}}$ [the prefactor of $T^2$ in Eq.~(\ref{Eq.(EP)})] vary between $\sim$$1.2\times10^{5}$ to $\sim$$2.2\times10^{5}$\,K$^{-2}$\,s$^{-1}$, which are two orders of magnitudes less than the values of $\tilde{A}$. Thus the \emph{e}-ph scattering rate is negligibly weak in our 3D In$_2$O$_3$ films.

In the framework of free-electron model, one can deduce the \emph{e}-ph relaxation rate $1/\tau_{e\text{-}t,{\text{ph}}}\propto n$ from Eq.~(\ref{Eq.(EP)}), where $n$ is the carrier concentration. However, Eq.~(\ref{Eq.(EE)}) predicates $1/\tau_{ee}^L \propto n^{-2/3}$ and $1/\tau_{ee}^S \propto n^{-4/3}$, where $\tau_{ee}^L$ and $\tau_{ee}^S$ represent the large- and small-energy-transfer \emph{e}-\emph{e} relaxation time, respectively. For the In$_2$O$_3$ films, the carrier concentration is around $\sim$$5\times 10^{19}$\,cm$^{-3}$, which is $\sim$3 to $\sim$4 orders of magnitudes less than that of typical metals. Hence the \emph{e}-\emph{e} scattering rate could be much greater than the \emph{e}-ph scattering rate in this low carrier concentration compound. The $1/\tau_\varphi(T)$ data were then least-squares fitted to the following equation:
\begin{equation}\label{Eq.(EEfit)}
\frac{1}{\tau}=\frac{1}{\tau_0}+A_{ee}^{S}T^{3/2}+A_{ee}^{L}T^2,
\end{equation}
where the second and third terms on the right-hand side of Eq.~(\ref{Eq.(EEfit)}) stand for the small- and large-energy-transfer\emph{e}-\emph{e} scattering terms, respectively. The solid curves in Fig.~\ref{FIGDephase} are the theoretical predication of Eq.~(\ref{Eq.(EEfit)}). Inspection of Fig.~\ref{FIGDephase} indicates that the experimental dephasing rate can be well described by Eq.~(\ref{Eq.(EEfit)}) in the whole measuring temperature range. The obtained values of $A_{ee}^{S}$ and $A_{ee}^{L}$, as well as $1/\tau_{\varphi}^{0}$, are listed in Table~\ref{TableLi}. Using free-electron-like model, one can easily obtain the theoretical values of $A_{ee}^{S}$ and $A_{ee}^{L}$ [see Eq.~(\ref{Eq.(EE)})] , denoted as $(A_{ee}^{S})^{\rm th}$ and $(A_{ee}^{L})^{\rm th}$ and also listed in Table~\ref{TableLi}, respectively. For most of the films, the values of $A_{ee}^{S}$ are nearly equal to the theoretical ones, except for films Nos. 2, 3 and 4. Even for the three films, the values of $A_{ee}^S$ and $(A_{ee}^{S})^{\rm th}$ agree to within a factor of $\sim$2 or smaller. Also, our experimental values of $A_{ee}^L$ are within a factor of $\sim$1.6 of $(A_{ee}^{L})^{\rm th}$. These levels of agreement are satisfactory. Thus, both the small- and large-energy-transfer \emph{e}-\emph{e} scattering processes govern the dephasing in these 3D In$_2$O$_3$ films. Inspection the experimental values of $A_{ee}^S$ and $A_{ee}^L$ indicates that the large-energy-transfer \emph{e}-\emph{e} dephasing rate is about one-half of that of the small-energy-transfer one even at 5\,K for each film.

\begin{figure}[htp]
\begin{center}
\includegraphics[scale=1]{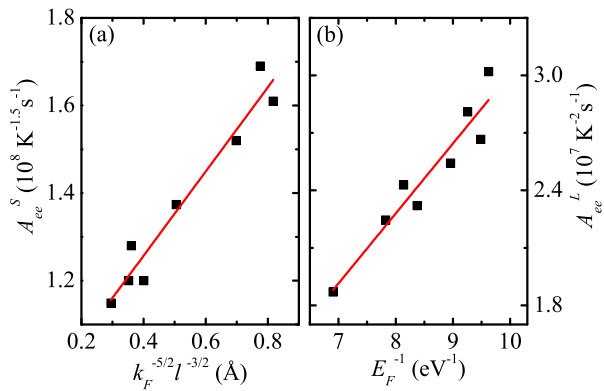}
\caption{(Color online)  (a) The small-energy-transfer $e$-$e$ scattering strength  $A_{ee}^{S}$ as a function of $k_F^{-5/2}\ell^{-3/2}$. (b) The large-energy-transfer $e$-$e$ scattering strength $A^L_{ee}$ as a function of $E_F^{-1}$. The solid lines are the linear fits to the experimental data.}
\label{FIGEDephase}
\end{center}
\end{figure}

According to Eq.~(\ref{Eq.(EE)}), the parameters $A_{ee}^S$ and $A_{ee}^L$ in Eq.~(\ref{Eq.(EEfit)}) should obey $A_{ee}^S\propto E_F^{-1/2} (k_F\ell)^{-3/2}$ and $A_{ee}^L\propto E_F^{-1}$, respectively. (In free-electron-like approximation, $A_{ee}^S$ should obey $A_{ee}^S\propto k_F^{-5/2}\ell^{-3/2}$.)  Figures~\ref{FIGEDephase}(a) and ~\ref{FIGEDephase}(b) show $A_{ee}^S$ variation with $k_F^{-5/2}\ell^{-3/2}$ and $A_{ee}^L$ as a function of $E_F^{-1}$, respectively. Here the values of $E_F$, $k_F$ and $\ell$ are also determined from the
free-electron-like model. As expected, the $A_{ee}^L\propto E_F^{-1}$ and $A_{ee}^S\propto k_F^{-5/2}\ell^{-3/2}$ rules are observed in Fig.~\ref{FIGEDephase}.
According to Eq.~(\ref{Eq.(EE)}), the slopes of $A_{ee}^S$-$k_F^{-5/2}\ell^{-3/2}$ and $A_{ee}^L$-$E_F^{-1}$ curves should be 3.41$\times10^{27}$\,m$^{-1}$\,K$^{-3/2}$\,s$^{-1}$ and 7.10$\times10^{-13}$\,J\,K$^{-2}$\,s$^{-1}$, respectively. The least-squares fits results indicate the slopes being $9.62\times 10^{26}$\,m$^{-1}$\,K$^{-3/2}$\,s$^{-1}$ and 5.84$\times10^{-13}$\,J\,K$^{-2}$s$^{-1}$, respectively. The experimental value of the slope of $A_{ee}^L$-$E_F^{-1}$ curve is $\sim$18\% smaller than the theoretical one. The consistency between the two values is extremely good. While the experimental and theoretical values of the slope of  $A_{ee}^S$-$k_F^{-5/2}\ell^{-3/2}$ curve agree to within a factor of $\sim$3. This level of agreement is also satisfactory.

As mentioned above, the carrier concentrations of the In$_2$O$_3$ films are $\sim$$5\times10^{19}$\,cm$^{-3}$, which are 3-4 orders of magnitude lower than that in typical metals. Hence the \emph{e}-ph relaxation rates of the In$_2$O$_3$ films are greatly suppressed and the \emph{e}-\emph{e} scattering rates are much enhanced. The $k_F\ell$ value of the In$_2$O$_3$ film is about twice as large as that in ITO films used in Ref.~\onlinecite{ref14}, while the carrier concentration of the former is about one fourth of that of the latter. Since the Fermi energy $E_F\propto n^{2/3}$, the small-energy-transfer \emph{e}-\emph{e} scattering strength $A_{ee}^S$ of the In$_2$O$_3$ film would be much less than that of the ITO film used in Ref.~\onlinecite{ref14}, while the large-energy-transfer \emph{e}-\emph{e} scattering strength $A_{ee}^L$ in the former would be great than that in the latter (see Eq.~(\ref{Eq.(EE)})). That is why the electron dephasing process in ITO films is only governed by the small-energy-transfer \emph{e}-\emph{e} scattering and both the large- and small-energy-transfer \emph{e}-\emph{e} scattering process have to be considered in In$_2$O$_3$ films. In a word, theses characteristics of relative large $k_F\ell$ and low carrier concentrations of In$_2$O$_3$ thick films give us opportunity to simultaneously observe small- and large-energy-transfer electron-electron scattering in 3D disordered conductors, and for the first time fully demonstrate the 3D $e$-$e$ scattering rate deduced 40 years ago.

\section{Conclusion}
We have studied the temperature behavior of resistivity and the electron dephasing rate in low temperature regime in In$_2$O$_3$ thick films. The $\rho(T)$ data  obey Bloch-Gr\"{u}neisen law from 300 down to $\sim$100\,K, indicating the films possess degenerate semiconductor characteristic in electrical transport properties. The \emph{e}-ph scattering rate is negligibly week though the In$_2$O$_3$ films are 3D with regard to WL effect. On the contrary, the \emph{e}-\emph{e} inelastic scattering govern the low temperature dephasing processes. In addition, besides the small-energy-transfer \emph{e}-\emph{e} scattering, the large-energy-transfer
\emph{e}-\emph{e} scattering also has significant contribution to the total electron relaxation rate. Our results also quantitatively demonstrate the validity of the theoretical predications of both small- and large-energy-transfer \emph{e}-\emph{e} scattering rates in experiment.

\begin{acknowledgments}
This work was supported by the
National Natural Science Foundation of China (NSFC)
through Grant No. 11174216, Research Fund for the
Doctoral Program of Higher Education through Grant No.
20120032110065.
\end{acknowledgments}


\begin{thebibliography}{00}
\bibitem{note1} The value of $q_T\ell$ is evaluated through  $q_T\ell\approx k_BT\ell/\hbar u_t$, where $u_t$ is the transverse sound velocity.
\end{thebibliography}
\end{document}